# Spectrum of density fluctuations in Brans–Dicke chaotic inflation

Mikel Susperregi[1]
*Theoretical Physics, University of Oxford,*
*1 Keble Road, Oxford OX1 3NP, United Kingdom.*

## Abstract

In the context of Brans–Dicke theories, eternal inflation is described in such a way that the evolution of the inflaton field is determined by the value of the Planck mass in different regions of the universe. The Planck mass is given by the values of the Brans–Dicke field, which is coupled to the scalar curvature in the Lagrangian. We first calculate the joint probability distributions of the inflaton and Brans–Dicke fields, in order to compute the 3–volume ratios of homogeneous regions with arbitrary values of the fields still undergoing inflation with respect to thermalized regions. From these volume ratios one is able to extract information on the values of the fields measured by a typical observer for a given potential and, in particular, the typical value of the Planck mass at the end of inflation. In this paper, we investigate volume ratios using a regularization procedure suggested by Vilenkin, and the results are applied to powerlaw and double–well potentials. The spectrum of density fluctuations is calculated for generic potentials, and we discuss the likelihood of various scenarios that could tell us whether our region of the universe is typical or untypical depending on very general bounds on the evolution of the Brans–Dicke field.

---

[1] *Electronic address*: mps@thphys.ox.ac.uk

# 1 Introduction

It is widely accepted within the standard framework of quantum cosmology that constants of nature take different values in different regions of the universe (Linde 1986,1989,1990a; Coleman 1988; Vilenkin 1995a,b; Garcia–Bellido & Linde 1995). One of the chief goals of quantum cosmology is to predict the probability distributions for these constants (e.g. the gravitational constant $G$, the cosmological constant, the density parameter $\Omega$, etc) from the wave function of the universe. By predicting the values of the constants of nature for a "typical" region of the universe from first principles, one would in principle have a means to verify whether we do indeed inhabit a typical region of the universe or not by comparison with the observational data. This is however a difficult undertaking, and it is customary to adopt the reverse approach, i.e. that of assuming that it is likely that we do indeed inhabit a typical region of the universe and, therefore, our predictions for the values of the constants of nature *should* be consistent with the data in the observable universe. This is a form of "principle of mediocrity" that has been enunciated in terms of anthropic arguments in the literature (Carter 1983; Barrow & Tipler 1986; Rees 1993; Albrecht 1995).

In models of eternal inflation, it is non–trivial to derive the probability distributions for the fields directly from the wave function of the universe (Linde 1995; Vilenkin 1995b), which is based on the Euclidean approximation to the sum over histories. This approach is not as yet entirely understood and yields unrenormalizable probability distributions for the fields that are difficult to interpret (Hartle & Hawking 1983; Vilenkin 1984; Hawking 1987; Linde 1984,1995). Alternatively, the Fokker–Planck or diffusion equation enables us to compute these probability distributions based on the idea that the inflaton field evolves stochastically and thus takes different values within a given region of the universe in a chaotic manner (see e.g. Starobinsky 1984,1986; Linde 1986). The evolution of the field is governed by Langevin–like equations of motion, where the classical motion of the field is perturbed by its Brownian motion in the form of quantum jumps around the classical slow–roll solution. As a result, regions where the field rolls down the potential below the end–of–inflation boundary are thermalized, whereas those in which inflation still takes place are subdivided in further regions where the field takes different values, some of which thermalize and others continue inflating (Linde 1986,1987; Goncharov *et al.* 1987; Garcia–Bellido 1994). In this scenario, provided one starts out from sufficiently homogeneous initial conditions, at any given time there will be a fraction of the total volume of the universe that has thermalized, $\mathcal{V}_*$, and the rest still undergoes inflation. Chaotic inflation predicts that this process is eternal and it can be implemented for a wide class of potentials (see e.g. Vilenkin 1983; Linde 1986). Also, it is found (Linde & Mezhlumian 1993) that the fraction of the physical volume where the inflaton field has a certain value $\sigma$ attains a constant value in the $t \to \infty$ limit. This is what we will refer to hereafter as the concept of "global stationarity". The implications of global stationarity for structure formation are strong as we shall see in the following.

In order to compute the probability distributions of the constants of nature or the spectrum of density fluctuations in different regions of the universe, it is necessary to compare the physical volumes of the hypersurfaces on which the fields take certain constant values. Provided that a sufficient amount of time has ellapsed since the onset of inflation, so that global stationarity can be assumed, the fraction of the physical volume of the universe where a constant of nature has an arbitrary value will be constant. In the case of the gravitational constant $G$ for example, we can investigate the spectrum of fluctuations in thermalized regions of the universe where $G$ takes different values. The value of $G$ is related to the Planck mass, and we expect this to be slowly varying in different regions of the universe, as described by a Brans–Dicke (BD) theory of gravity (see e.g. Brans & Dicke 1961; Brans 1962; Dicke 1962). Regions of the universe where the value of $G$ is greater will experience a quicker process of galaxy formation and cosmic structures will be clumpier for the same value of the density parameter $\Omega$. Inflation is brought in this scenario by models of extended inflation (see e.g. La & Steinhardt 1989; Linde 1990b) as well as inflationary scenarios that incorporate in the Lagrangian a BD gravity theory (Linde 1990a,1994a; Laycock & Liddle 1994).

In chaotic inflation with BD gravity (Linde 1990b; Garcia–Bellido 1994; Garcia–Bellido & Linde 1995) the value of the Planck mass before the onset of inflation places an upper bound on the values of the inflaton field, above which metric fluctuations become too large for inflation to occur. On the other hand, the Planck mass varies slowly as the inflaton field rolls down the potential, and its value remains constant at the end of inflation. Therefore, the value of the Planck mass at the end of inflation dictates the formation of structure and quantum physics in thermalized regions. By virtue of global stationarity, one can calculate the probability distribution of $G$ and the spectrum of density perturbations in those regions in a time–independent manner. In this paper we derive the joint probability distributions of the inflaton and BD fields within this scenario, therefore predicting the likeliest values of the Planck mass at the end of inflation given the initial conditions, and the spectrum of density fluctuations is computed therefrom.

In order to calculate the spectrum of density fluctuations in different regions we have mentioned that we need to know the volume ratio of these regions. A way to proceed is to choose a foliation where the time variable is the synchronous time and calculate the 3-volumes at constant time. We construct the ratio

$$r = \frac{\mathcal{V}(t,\sigma_0)}{\mathcal{V}_*(t)}, \qquad (1)$$

which gives us the relative probability of the inflaton field $\sigma$ taking values in the range $\sigma_0, \sigma_0+d\sigma$ with respect to the thermalized regions ($\sigma \lesssim \sigma_*$) at time $t$. It is expected that this ratio will converge to a constant value in the limit $t \to \infty$. It has been pointed out however (Garcia–Bellido et al. 1994; Linde et al. 1994,1995) that the calculation of the probabilities is extremely sensitive to the choice of time parametrization, and different choices of $t$ yield to inconsistent results in the asymptotic limit.

Recently, a method was proposed (Vilenkin 1995a) whereby one can investigate the ratios of physical 3–volumes of different regions of the universe in a way that is very insensitive to the choice of time parametrization. The method consists on computing the 3–volumes of hypersurfaces $\sigma = \mathrm{const}$, not at $t = \mathrm{const}$ but over all spacetime. Thus, one construct ratios similar to

(1),
$$r = \frac{\mathcal{V}(\sigma_1)}{\mathcal{V}_*}, \qquad (2)$$

where the volumes are integrated for all times. Certainly the individual volumes are infinite, but due to global stationarity, the ratio of any two volumes $\mathcal{V}(\sigma_1)$ and $\mathcal{V}(\sigma_2)$ remains finite. The procedure to calculate these ratios consists on a regularization proposal that is summarized in Section 3.

From global stationarity, we have that the comoving probability distribution of the inflaton field reaches an asymptotic regime,
$$\mathcal{P}(t \to \infty, \sigma) = \mathcal{P}_S(\sigma), \qquad (3)$$
which is related to the 3–volumes $\mathcal{V}$ via
$$\mathcal{P}_S(\sigma) = \frac{\mathcal{V}(\sigma)}{\mathcal{V}_T}, \qquad (4)$$
where
$$\mathcal{V}_T \equiv \int \mathrm{d}\sigma\, \mathcal{V}(\sigma). \qquad (5)$$

By calculating the probability distribution of the inflaton field, the regularization method can be used to compute the spectrum of density fluctuations in a way that is insensitive of the choice of time variable.

In this paper, we investigate the application of the regularization procedure of Vilenkin (1995a) to models of chaotic inflation in a BD theory of gravity, and we derive the spectrum of density fluctuations. The article is organized as follows: in Section 2 we describe the fundamentals of eternal inflation in a gravity theory that is of the BD type for two different potentials; in Section 3 the regularization procedure for comparing infinite volumes of constant–field hypersurfaces is summarized for an inflation–only model (BD field constant); in Section 4 we calculate the probability distributions in a generic BD chaotic inflation we apply the tools of Section 3 for the derivation of the volume ratios given by these probability distributions; in Section 5 we derive the amplitudes of the quantum fluctuations of the fields measured by a typical observer and we use these results to calculate the spectrum of density fluctuations; finally in Section 6 we discuss the results and possible observational predictions.

## 2  Eternal inflation in Brans–Dicke gravity

The BD action in chaotic inflation is given by (e.g. Barrow & Maeda 1991; Liddle & Wands 1992; Garcia–Bellido *et al.* 1994):
$$\mathcal{S} = \int \mathrm{d}^4 x\, \sqrt{-g}\, \Big[\frac{1}{8\omega}\phi^2 R - \frac{1}{2}(\partial\phi)^2 - \frac{1}{2}(\partial\sigma)^2 - V(\sigma)\Big], \qquad (6)$$

where $R$ is the spacetime curvature and $V(\sigma)$ is an arbitrary inflaton potential. The Planck mass is related to the BD field $\phi$ via

$$G^{-2} = M_P^2(\phi) = \frac{2\pi}{\omega}\phi^2, \tag{7}$$

where the coupling constant $\omega$ satisfies the observational bound $\omega \gtrsim 500$ (Reasenberg et al. 1979; Accetta et al. 1990; Casas et al. 1992; Will 1993) and in the limit $\omega \to \infty$ one recovers GR, where the value of $G$ is constant throughout. In the context of string theory, the coupling $\omega(\phi)$ becomes a dynamical variable for the so-called dilaton field $\phi$ (see e.g. Bergmann 1968; Nordtvedt 1970; Wagoner 1970; Damour & Nordtvedt 1993; Lidsey 1996). The BD field is a particular case of dilaton field for which the coupling $\omega$ with matter is constant, as we shall assume hereafter.

The Planck boundary or beginning–of–inflation boundary is given by the curve

$$V(\sigma) = M_P^4(\phi), \tag{8}$$

that marks the boundary were metric fluctuations begin to be large, and hence, we do not allow the inflaton field to take values for which the potential would be above this boundary. In the case of arbitrarily large values of $\phi$, a larger range of values of $\sigma$ is permitted, and thus one may encounter the situation where the fields diverge simultaneously while remaining within the curve (8). On the other hand, the end of inflation is marked by the condition

$$\frac{1}{2}\dot\sigma^2 + \frac{1}{2}\dot\phi^2 \approx V(\sigma), \tag{9}$$

which, like the Planck boundary (8), is also in principle unbounded.

The equations of motion in an FRW background become:

$$\left(\mathcal{D}^2 + \frac{1}{4\omega}R\right)\phi = 0, \tag{10}$$

$$\mathcal{D}^2\sigma = -V'(\sigma), \tag{11}$$

$$H^2 + \frac{k}{a^2} + 2H\frac{\dot\phi}{\phi} = \frac{4\omega}{3\phi^2}\left[\frac{1}{2}\dot\phi^2 + \frac{1}{2}\dot\sigma^2 + V(\sigma)\right], \tag{12}$$

where the curvature $k = 0, \pm 1$ and the differential operator $\mathcal{D}$ is defined

$$\mathcal{D}^2 \equiv \partial_t^2 + 3H\,\partial_t - \frac{k^2}{a^2}. \tag{13}$$

In the slow–roll approximation, i.e. during inflation, where $\ddot\phi \ll H\dot\phi \ll H^2\phi$, $\dot\sigma^2 + \dot\phi^2 \ll 2V(\sigma)$ and $V''(\sigma), k^2\,a^{-2} \ll H^2$, these equations read

$$\frac{\dot\phi}{\phi} = \frac{H}{\omega}, \tag{14}$$

$$\dot\sigma = -\frac{1}{3H}V'(\sigma), \tag{15}$$

$$H^2 = \frac{4\omega}{3\phi^2} V, \qquad (16)$$

and the curvature scalar is given by $R = -12H^2$. Equations (14)–(16) permit us to write the end–of–inflation boundary (9) in the form

$$\dot{\phi}^2 = 8(3\omega - 2) \left(\frac{V}{V'}\right)^2. \qquad (17)$$

Thereupon, in the limit of large $\omega$, we obtain that the Planck mass at the end of inflation is given by

$$M_P^2 = 48\pi \left(\frac{V}{V'}\right)^2. \qquad (18)$$

In our own observable part of the universe, $M_P \sim 10^{19}$ GeV, and therefore,

$$\left|\frac{V(\sigma_*)}{V'(\sigma_*)}\right| \sim 10^{18} \text{GeV}, \qquad (19)$$

where $\sigma_*$ denotes the value of the inflaton at the end of inflation. It is easy to show also from (14)–(16) that the following general conservation law holds:

$$\frac{\mathrm{d}}{\mathrm{d}t}\left[\dot{\phi}^2 + 8\int \mathrm{d}\sigma \frac{V(\sigma)}{V'(\sigma)}\right] = 0. \qquad (20)$$

Notice also that the evolution of the inflaton field is determined by the BD field through the Hubble parameter in the damping term on the RHS of (15). This effect will be greater the smaller $H$ is, which from (14) one can see occurs for slow variations of $\log \phi$. In the following we discuss the dynamics of this scenario for two inflaton potentials.

## 2.1 Model I: Power–law potential

In this case we consider potentials of the type

$$V(\sigma) = \frac{\lambda}{2n} \sigma^{2n}, \qquad (21)$$

where $n$ is an integer an the coupling $\lambda$ is typically of order unity. The Planck boundary is then given by

$$\phi_P^2 = \left(\frac{\omega}{2\pi}\right)\left(\frac{\lambda}{2n}\right)^{1/2} \sigma_0^n. \qquad (22)$$

Thus, we require that initially $\phi_0 > \phi_P$ given an arbitrary $\sigma_0$. Hence, at the initial time, provided the latter condition is satisfied, one may find the situation where both $\phi_0$ and $\sigma_0$ are very large at the onset of inflation but still the inflaton field lies under the Planck boundary. The equations of motion in the slow–roll regime become

$$\frac{\dot{\phi}}{\phi} = \frac{H}{\omega}, \qquad (23)$$

$$\frac{\dot{\sigma}}{\sigma} = -\frac{n}{2}\frac{H}{\omega}\frac{\phi^2}{\sigma^2}, \tag{24}$$

$$H^2 = \frac{4\omega}{3\phi^2}\frac{\lambda}{2n}\sigma^{2n}. \tag{25}$$

The end–of–inflation boundary is given by the curve

$$\phi_*^2 = \frac{2}{n^2}\sigma_*^2\,(3\omega - 2), \tag{26}$$

where the star–subscript quantities denote the values of the fields at the end of inflation, and from (23)(24) again we derive a conserved quantity:

$$\frac{\mathrm{d}}{\mathrm{d}t}\left(\phi^2 + \frac{2}{n}\sigma^2\right) = 0, \tag{27}$$

which tells us that the classical motion of the fields in the slow–roll regime is along a ellipse on the $(\phi, \sigma)$ plane as is illustrated in Fig. 1. It is apparent from Fig. 1 that, given some initial conditions $(\sigma_0, \phi_0)$, as the inflaton field rolls down the potential during inflation by action of its classical drift, the BD field increases in value monotonically and its rate of growth is weakly dependent on $n$. Therefore, the Planck mass at the end of inflation is always greater than its initial value. The three figures correspond to potentials $n < 2$, $n = 2$ and $n > 2$. In all cases, the end–of–inflation curve is a straight line, whereas the Planck boundary is strongly dependent on the value of $n$. Only in those cases where $n > 2$ do the Planck boundary and the end–of–inflation boundary intersect, and the enclosed finite area is the only region where inflation can take place. This effectively sets an upper bound for the values of the fields at the initial times and, ultimately, for the value of the Planck mass at the end of inflation. In the other two cases the initial conditions are effectively unbounded, provided that they lie in the region above (22).

Conservation of the quantity $\phi^2 + (2/n)\sigma^2$ at the initial and end–of–inflation times yields

$$\phi_0^2 + \frac{2}{n}\sigma_0^2 = \frac{2}{n}\sigma_*^2\left[1 + \frac{1}{n}(3\omega - 2)\right]. \tag{28}$$

Hence,

$$\sigma_*^2 = \left(\sigma_0^2 + \frac{n}{2}\phi_0^2\right)\left[1 + \frac{1}{n}(3\omega - 2)\right]^{-1}, \tag{29}$$

$$\phi_*^2 = \frac{2}{n^2}(3\omega - 2)\left(\sigma_0^2 + \frac{n}{2}\phi_0^2\right)\left[1 + \frac{1}{n}(3\omega - 2)\right]^{-1}. \tag{30}$$

If $\phi_0 \gg \sigma_0$, then the fields at the end–of–inflation boundary are fully determined by the BD field. In chaotic inflation, the initial situation is that the inflation field starts out from values close to the Planck boundary and rolls down the potential towards smaller values. This implies, as pointed out above, that the BD field grows and therefore the Planck mass at the end of inflation is larger than its initial value. However, in the case of $\phi_0 \gg \sigma_0$, and in the limit $\omega \to \infty$, the Planck mass remains practically constant as it is easy to show from (29)(30). We get

$$\sigma_*^2 \approx \frac{n^2}{6\omega}\phi_0^2, \tag{31}$$

$$\phi_*^2 \approx \phi_0^2. \tag{32}$$

On the other hand, from (18), we get that if a powerlaw potential is a good approximation to the inflaton potential in our observable universe, then

$$\left|\frac{\sigma_*}{2n}\right| \sim 10^{18} \text{ GeV}, \tag{33}$$

or equivalently,

$$\left|\frac{\phi_0}{\omega^{1/2}}\right| \sim 10^{19} \text{ GeV}, \tag{34}$$

which requires a large value of $\phi_0$ for most conservative estimates of $\omega$.

## 2.2 Model II: Double–well potential

Let us consider the case of a symmetric double–well inflaton potential such as:

$$V(\sigma) = \frac{1}{4}m^2 \, (\sigma^2 - \eta^2)^2. \tag{35}$$

The Planck boundary is then given by the curve

$$\left(\frac{m\omega}{4\pi}\right) |\sigma^2 - \eta^2| = \phi^2. \tag{36}$$

Given $\sigma \approx 0$ initially, the Planck boundary is precisely on top of the hill at $\sigma = 0$ if

$$\phi^2 = \phi_P^2 \equiv \frac{\omega m}{4\pi}\eta^2, \tag{37}$$

with a corresponding Planck mass $M_P^2 = (m/2)\,\eta^2$. Thus, if $\phi(t \to 0) > \phi_P$ the Planck boundary will be located above the maximum of the potential at $\sigma = 0$, and if $\phi(t \to 0) < \phi_P$ it will cut through the slope of the potential. In the latter case, the allowed initial value of $\sigma$ is greater than a threshold value

$$|\sigma_0| > \left[\eta^2 - \left(\frac{4\pi}{m\omega}\right)\phi_0^2\right]^{1/2}, \tag{38}$$

so that $\sigma$ will lie under the Planck boundary. The subindex 0 denotes the values of the fields at $t = 0$. The equations of motion in the slow–roll approximation are:

$$\frac{\dot{\sigma}^2}{\sigma^2} = \left(\frac{m^2}{3\omega}\right)\phi^2, \tag{39}$$

$$\dot{\phi}^2 = \left(\frac{m^2}{3\omega}\right)(\sigma^2 - \eta^2)^2, \tag{40}$$

$$H^2 = \frac{m^2\omega}{3\phi^2}(\sigma^2 - \eta^2)^2, \tag{41}$$

and therefore the end–of–inflation boundary is given by

$$\phi_*^2 \left(\frac{2}{3\omega - 2}\right) = \frac{(\sigma_*^2 - \eta^2)^2}{\sigma_*^2}. \tag{42}$$

Hence, the inversion of (42) gives us the value of the inflaton at the end of inflation,

$$\sigma_* = \frac{1}{2}\left(\frac{2}{3\omega - 2}\right)^{1/2} \phi_* \left\{-1 + \left[1 + \frac{2\eta^2}{\phi_*^2}(3\omega - 2)\right]^{1/2}\right\}, \tag{43}$$

and the Planck mass at this epoch is

$$M_P^2(\phi_*) = \frac{\pi}{\omega}(3\omega - 2)\frac{(\sigma_*^2 - \eta^2)^2}{\sigma_*^2}. \tag{44}$$

Thus, in the limit $\omega \to \infty$, assuming a double–well inflaton potential is a good approximation in the observable universe, one gets

$$\frac{|\sigma_*^2 - \eta^2|}{\sigma_*} \sim 10^{18}\,\text{GeV} \tag{45}$$

at the end of inflation and, therefore, a very large value of $\eta$ is required such that $|\sigma_* - \eta| \ll \eta$. This in itself implies that $\eta \gg 10^{18}$ GeV. Therefore the potential is very flat and inflation terminates a long while before $\sigma$ has reached the minimum of the potential. From (39)(40), the conservation law (20) takes the form

$$\frac{\mathrm{d}}{\mathrm{d}t}\left(\phi^2 + \sigma^2 - \eta^2 \log \sigma^2\right) = 0 \tag{46}$$

and therefore the motion of the fields is confined to a circle in the $(\phi, u)$ plane, where $u$ is given by

$$u^2 = \sigma^2 - \eta^2 \log \sigma^2. \tag{47}$$

Let us consider the situation where the inflaton field is initially small, $\sigma_0 \approx \epsilon$. The conservation law (46) enables us to describe the evolution of the BD field as the inflaton rolls–down the potential towards the end–of–inflation boundary. Hence

$$\phi_0^2 - \eta^2 \log \epsilon^2 \approx \phi_*^2 + \sigma_*^2 - \eta^2 \log \sigma_*^2, \tag{48}$$

where we have neglected a $O(\epsilon^2)$ term. By imposing the bounds $\sigma_*^2 > 0$ and $\sigma_*^2 < \eta^2$, we get

$$\eta^2\left[1 + \log\left(\frac{\epsilon^2}{\sigma_*^2}\right)\right] > \phi_0^2 - \phi_*^2 > \eta^2 \log\left(\frac{\epsilon^2}{\sigma_*^2}\right). \tag{49}$$

Both bounds on the left and right of (49) diverge in the limit $\epsilon \to 0$, and thus

$$\phi_0^2 - \phi_*^2 \sim \log \epsilon \to -\infty \qquad \epsilon \to 0. \tag{50}$$

Therefore, if the inflaton field starts from $\sigma \approx 0$ at the initial time, the BD field will grow indefinitely in the slow–roll regime regardless of its initial value $\phi_0$:

$$\phi_*^2 \sim \phi_0^2 + \eta^2 \log\left(\frac{\sigma_*^2}{\sigma_0^2}\right). \tag{51}$$

If however, (1) $\sigma_0 > 0$ (due to the Planck boundary being on the slope of the potential for example, i.e. $\phi_0^2 < \frac{m\omega}{4\pi}\eta^2$), or (2) $\sigma_*$ is unbounded, then $\phi$ will remain thoroughly finite and thus, the resulting Planck mass at the end of inflation will be finite. We can obtain a Planck mass of $\sim 10^{19}$ GeV in our observable universe in the limit $\omega \to \infty$ provided that the ratio $\frac{\phi_*^2}{\omega}$ is large. This will require that either the initial $\phi_0$ is very large, or alternatively, that $\sigma_0$ be close to zero. In either case, the possibility of having the Planck boundary across the slope of the potential is ruled out because this would imply a small value of $\phi_0$, and also a large $\sigma_0$ since, as we have concluded above, the potential is very flat around its maximum and minima due to the large value of $\eta$. Thus, a further constraint arises on the parameter $m$ of the potential:

$$M_P(\phi_*) > \left(\frac{\omega m}{4\pi}\right)^{1/2} \eta \gg m^{1/2}\, 10^{19}\ \text{GeV}, \qquad (52)$$

and therefore $m \ll 1$, where we have used $(\omega/4\pi)^{1/2} \gtrsim 10$. As is well known (see e.g. Linde 1990b), we need $m \sim 10^{-6} M_P$ in order to produce density perturbations over scales of astrophysical interest $\delta\rho/\rho \sim 10^{-5}$, which is consistent with the constraint (52).

## 3  Comparing volumes in an eternally inflating universe

In this section, I summarize the regularization procedure proposed in Vilenkin (1995a) (in an inflation–only scenario) for the case of a powerlaw potential (21). In the following section, these results will be extrapolated to the BD model, that we will use in order to calculate the spectrum of density fluctuations. At a sufficiently early time, we pick a hypersurface on which $\sigma \approx \sigma_0$, and the radius of the universe is normalized to unity at this epoch, $a = 1$. Assuming an eternally inflating scenario, one is able to choose this normalization at an arbitrary radius of the universe. Let us denote the 3–volume of this hypersurface $\mathcal{V}_0$. A number $N$ of observers located on this hypersurface will, after a given period of inflation, evolve into the hypersurface $\sigma \equiv \sigma_*$, which is the value of the field at the end–of–inflation boundary at the slope of the potential. The initial volume per observer on the $\sigma \approx \sigma_0$ hypersurface is given by $\mathcal{V}_i(0) = \mathcal{V}_0/N$ and after a time $a_{*i}$, the $i$–th observer will be located on the hypersurface $\sigma \approx \sigma_*$ and its volume will grow by a factor of $a_{*i}^3$. Therefore, the total 3–volume of the hypersurface $\sigma = \sigma_*$ is given by

$$\mathcal{V}_* = \frac{\mathcal{V}_0}{N} \sum_i a_{*i}^3, \qquad (53)$$

where the sum is over all the observers that reach $\sigma_*$. In the limit $N \to \infty$, the sum in (53) diverges due to the fact that for any arbitrarily large time there will always be regions undergoing inflation, and thus there are $a_{*i}$ terms growing without limit. The regularization procedure that we will use consists in setting a cutoff in the sum (53) for a small fraction $\epsilon$ of regions with the largest values of $a_{*i}$. Therefore, one can write $\mathcal{V}_*$ in terms of the parameter $\epsilon$ and thus compute the ratio of the 3–volumes of any $\sigma = \text{const}$ hypersurfaces in the limit $\epsilon \to 0$. The success of this procedure lies in that the regularized volumes of the hypersurfaces of constant $\sigma$ is calculated over all spacetime and it is therefore independent on the choice of spacetime foliation.

## 3.1 Volume ratios

The comoving probability distribution $\mathcal{P}_c(\sigma, t)$ of the inflaton field $\sigma$ satisfies the continuity equation

$$\partial_t \mathcal{P}_c = -\partial_\sigma J, \tag{54}$$

where the probability current

$$J = \frac{1}{8\pi^2} H^{\frac{\alpha}{2}+1} \partial_\sigma (H^{\frac{\alpha}{2}+1} \mathcal{P}_c) - \frac{1}{4\pi} H^{\alpha-1} H' \mathcal{P}_c \tag{55}$$

and the parameter $\alpha$ defines a time parametrization $t$ that is related to the proper time $\tau$ via the relation (see e.g. Mijić 1990,1991)

$$\mathrm{d}t = H^{1-\alpha}\,\mathrm{d}\tau, \tag{56}$$

where $H$ is the Hubble parameter. The flux $J$ tells us the fraction of the comoving volume at a given hypersurface of constant value of $\sigma$, such that $|J(\sigma,t)|\,\mathrm{d}t$ is the fraction of observers on such a hypersurface between $t$ and $t + \mathrm{d}t$. The synchronous time parametrization is recovered for the case of $\alpha = 0$, such that $a = \mathrm{e}^t$ and thus

$$\mathcal{V}_*(\text{regularized}) = \mathcal{V}_0 \left| \int_0^{t_c} J(\sigma_*, t)\, \mathrm{e}^{3t}\,\mathrm{d}t \right|, \tag{57}$$

where the cutoff time $t_c$ is given by

$$\left| \int_{t_c}^\infty J(\sigma_*, t)\,\mathrm{d}t \right| = \epsilon \left| \int_0^\infty J(\sigma_*, t)\,\mathrm{d}t \right|. \tag{58}$$

The cutoff parameter $\epsilon$ regulates what fraction of the late–time integrated probability flux is removed from the regularized volume. In the limit $\epsilon \to 0$, we have $\mathcal{V}_* \to \infty$. It is easy to show that in the regime where diffusion can be neglected, i.e. during the slow–roll regime

$$J \approx -\frac{1}{4\pi} \frac{H'}{H} \mathcal{P}_c, \tag{59}$$

the ratio of the 3–volume hypersurface $\sigma = \sigma_A$ (where $\sigma_A$ is an arbitrary value in the range $|\sigma_*| > |\sigma_A| \geq 0$) with respect to the thermalized volume $\mathcal{V}_*$ is

$$r \equiv \frac{\mathcal{V}_A}{\mathcal{V}_*} \sim \left[ \frac{\tau_\sigma(\sigma_A)}{\tau_\sigma(\sigma_*)} \right]^3, \tag{60}$$

where

$$\tau_\sigma(\sigma) \equiv \exp\left[ -4\pi \int_{\sigma_0}^\sigma \frac{H(\xi)}{H'(\xi)} \mathrm{d}\xi \right], \tag{61}$$

denotes the expansion factor during the classical slow–roll from the initial hypersurface $\sigma \approx \sigma_0$ and the final $\sigma$. The ratio (60) represents the relative probability of finding the inflaton field at a value $\sigma_A$ at an arbitrary spacetime point with respect to the thermalized or end–of–inflation value $\sigma_*$. By virtue of global stationarity, this relative probability is time–independent. We can rewrite (60) in the form

$$r = \exp\left[12\pi \int_{\sigma_A}^{\sigma_*} \frac{H(\xi)}{H'(\xi)} \, \mathrm{d}\xi\right]. \tag{62}$$

For the case of a double–well potential (35) we get

$$r_{DW} = \left|\frac{\sigma_A}{\sigma_*}\right|^{6\eta^2} \exp\left[-3\pi(\sigma_A^2 - \sigma_*^2)\right], \tag{63}$$

and similarly for a powerlaw potential (21)

$$r_{PL} = \exp\left[-\frac{6\pi}{n}(\sigma_A^2 - \sigma_*^2)\right]. \tag{64}$$

In the case of the double–well potential, as is well–known, the probability of the field for values away from $\sigma_*$ is exponentially suppressed, and rather less so in the case of the powerlaw potential, due to the factor preceding the exponential in (63). As shown in Vilenkin (1995a), the quantity $r$ is very insensitive to the choice of time parametrization and can be therefore used to good effect to compute the spectrum of density fluctuations, as will be discussed in §5. A mild dependence on the choice of parametrization is however present, and this is due to the no–diffusion approximation (59). On the other hand, it has been shown recently (Linde & Mezhlumian 1996) that, for regions undergoing inflation which are away from the non–diffusion regime, the dependence on the choice of time parametrization (as indeed the dependence on the choice of regularization procedure) can be rather strong, so the results become ambiguous as we depart from our slow–roll approximation. It is expected nonetheless, that in the regime where this approximation is valid (away from the Planck boundary and close to the end–of–inflation boundary), the results discussed in the remainder of this article will be consistent and insensitive to the choice of $\alpha$. We will discuss the performance of this approximation in the context of BD chaotic inflation in the following section.

# 4 Volume ratios in Brans–Dicke scenarios

## 4.1 Stationary probability distributions

In this section, we extend the results summarized above for inflation–only scenarios to BD eternal inflation. The probability distributions derived apply to generic potentials and we will investigate the application to the potentials (21)(35). In BD inflation, the comoving probability $\mathcal{P}_c(\phi, \sigma, t)$ of the inflaton and BD fields is given by the conservation equation (see e.g. Starobinsky 1986; Linde 1990b)

$$\partial_t \mathcal{P}_c = -\partial_\sigma J_\sigma - \partial_\phi J_\phi, \tag{65}$$

where the components of the probability current $\vec{J} \equiv (J_\sigma, J_\phi)$ are defined:

$$J_\sigma = -\frac{1}{8\pi^2} H^{\frac{\alpha}{2}+1} \, \partial_\sigma (H^{\frac{\alpha}{2}+1} \mathcal{P}_c) - \frac{1}{4\pi} H^{\alpha-1} \partial_\sigma H \, \mathcal{P}_c, \tag{66}$$

$$J_\phi = -\frac{1}{8\pi^2} H^{\frac{\alpha}{2}+1} \, \partial_\phi (H^{\frac{\alpha}{2}+1} \mathcal{P}_c) - \frac{1}{2\pi} H^{\alpha-1} \partial_\phi H \, \mathcal{P}_c, \tag{67}$$

and the Hubble parameter $H$ is given by (16). In the no–diffusion [slow–roll] regime that we shall adopt hereafter, i.e. on the slopes of the potential, the first term on the RHS of (66)(67) is negligible and thus

$$J_\sigma \approx -\frac{1}{4\pi} H^{\alpha-1} \partial_\sigma H \, \mathcal{P}_c, \tag{68}$$

$$J_\phi \approx -\frac{1}{2\pi} H^{\alpha-1} \partial_\phi H \, \mathcal{P}_c. \tag{69}$$

One can expand the probability distribution $\mathcal{P}_c(\phi, \sigma, t)$ in terms of eigenfunctions in order to transform (65) into an eigenvalue equation. We have

$$\mathcal{P}_c(\phi, \sigma, t) = \sum_{n=1}^{\infty} \psi_n(\phi, \sigma) \, e^{-\gamma_n t}, \tag{70}$$

where the eigenvalues $\gamma_i$ are ordered $\gamma_1 < \gamma_2 < \gamma_3 < \ldots$. In the limit $t \to \infty$ the dominant contribution is

$$\mathcal{P}(\phi, \sigma, t \to \infty) = \psi_1(\phi, \sigma) \, e^{-\gamma_1 t}, \tag{71}$$

and therefore (65) for $\psi_1$ reads

$$-4\pi \gamma_1 \psi_1 = \partial_\sigma \left[ H^{\alpha-1} \partial_\sigma H \, \psi_1 \right] + 2 \partial_\phi \left[ H^{\alpha-1} \partial_\phi H \, \psi_1 \right]. \tag{72}$$

In the following we calculate the volume ratios for the time parametrization given by $\alpha = 0$, which corresponds to $t = \log a$. As shown in Vilenkin (1995a), the regularization procedure summarized in §4 yields results that are insensitive to the choice of $\alpha$ in the slow–roll regime. The eigenvalue equation (72) then becomes

$$-4\pi \gamma_1 \, \psi_1 = \partial_\sigma (Z_\sigma^{-1} \psi_1) + 2 \, \partial_\phi (Z_\phi^{-1} \psi_1), \tag{73}$$

where we have defined for shorthand the quantities

$$Z_\sigma^{-1} \equiv H^{-1} \partial_\sigma H = \frac{1}{2} \frac{V'}{V}, \tag{74}$$

$$Z_\phi^{-1} \equiv H^{-1} \partial_\phi H = -\phi^{-1}, \tag{75}$$

which only depend on $\sigma$ and $\phi$ respectively. Given the initial conditions $(\phi_0, \sigma_0)$, then the solution of (73) is

$$\psi_1(\phi, \sigma) = 2 C_0 \, \phi \left( \frac{V}{V'} \right) \exp(\tau_\sigma + \tau_\phi), \tag{76}$$

where

$$\tau_\sigma(\sigma) \equiv -\frac{4\pi\gamma_1}{3} \int_{\sigma_0}^{\sigma} Z_\sigma(\zeta) \, d\zeta, \qquad (77)$$

$$\tau_\phi(\phi) \equiv -\frac{4\pi\gamma_1}{3} \int_{\phi_0}^{\phi} Z_\phi(\xi) \, d\xi, \qquad (78)$$

and $C_0$ is a normalization constant. Substituting (74)(75) in these integrals and making use of (20), we get

$$\tau_\sigma + \tau_\phi = \pi\gamma_1 \, (\phi^2 - \phi_0^2), \qquad (79)$$

regardless of the form of the potential. The term $e^{-\pi\gamma_1 \phi_0^2}$ in (76) can be absorbed in the normalization coefficient $C_0$ as will be assumed hereafter. Hence

$$\mathcal{P} \approx 2 C_0 \, \phi \left(\frac{V}{V'}\right) \, \exp(\pi\gamma_1 \, \phi^2 - \gamma_1 t). \qquad (80)$$

The parameters $C_0$ and $\gamma_1$ are determined by the normalization of the probability and the form of the potential. On the one hand, we have that at the initial time the probability is sharply peaked at values of the fields in the neighbourhood of the Planck boundary, thus:

$$\int_{\mathcal{N}} d\sigma \, d\phi \, \psi_1(\sigma, \phi) = 1, \qquad (81)$$

where $\mathcal{N}$ is a sufficiently broad region neighbouring the Planck boundary. In the case of the double–well potential, this is located on the top of the hill in the neighbourhood of $\sigma \approx 0$, whereas in the case of the powerlaw potential $\mathcal{N}$ is a region of large values of $\sigma$ and $\phi$, just below the Planck boundary. On the other hand, the value of $\gamma_1$ is computed by solving the eigenvalue equation for a given potential and magnitude of its coupling constant (see e.g. Linde & Mezhlumian 1993 and Linde 1994b for a similar calculation in the inflation–only scenario). Provided that the physical volume of the inflating regions grows as $a^{3t}$, from (71) we have that inflation will only be eternal if $\gamma_1 \leq 3$.

Within the slow–roll approximation, the solution (76) automatically satisfies the boundary conditions at the end–of–inflation boundary, namely the conditions of conservation of probability and probability flux, since the probability currents employed, (68)(69), are equal at either side of the end–of–inflation boundary, the diffusion terms having been neglected.

In the case of the double–well potential (35), we get

$$\mathcal{P}_c(\sigma, \phi, t) = C_0 \, \phi \left|\frac{\eta^2 - \sigma^2}{2\sigma}\right| \, \exp(\pi\gamma_1 \, \phi^2 - \gamma_1 t), \qquad (82)$$

and for the powerlaw potential (21),

$$\mathcal{P}_c(\sigma, \phi, t) = C_0 \, \phi \, \frac{\sigma}{n} \, \exp(\pi\gamma_1 \, \phi^2 - \gamma_1 t). \qquad (83)$$

In both cases, there is a greater probability for regions to continue inflating and for the BD field to take large values. Therefore, the Planck boundary is likelier to allow larger values of the inflaton field in the initial conditions, and the expectation value of the Planck mass is expected to peak around a large value. In the case of (82), the probability is higher for the inflaton to remain away from $\sigma^2 \approx \eta^2$, and thus, most regions have values of $\sigma$ on the slope of the potential and therefore continue inflating. In the case of (83), the inflaton field is likelier to take small values so that, by virtue of the conservation law (27), the BD field will grow as much as possible.

In order to calculate the probability distribution of the fields along the end–of–inflation boundary we substitute (17) in (76) and thus

$$\mathcal{P}_c\big|_{EoI} = C_0 \left[2(3\omega - 2)\right]^{-1/2} \phi^2 \, \exp(\pi\gamma_1 \, \phi^2 - \gamma_1 t), \tag{84}$$

which is a monotonically growing function in $\phi$ and its derivative only vanishes at $\phi = 0$. Therefore, the likeliest value of $\phi$ in any scenario is the highest one allowed by the form of the potential. The possibility of the so–called "run–away" solutions (i.e. solutions where both fields grow without limit along the Planck boundary or below it) remains viable in this scenario, unless the potential is such that the Planck boundary intersects the end–of–inflation boundary, like in the example of powerlaw $n > 2$ that was discussed in §3. It could be conjectured that if $M_P$ in our own neighbourhood has a typical value with respect to other regions in the universe, then the real inflaton potential will be such that such an intersection of boundaries ought to take place and hence the calculation of $\mathcal{P}_c$ would produce a distribution that does not allow runaway solutions.

## 4.2 Physical volumes of thermalized and inflating regions

The total 3–volume $\mathcal{V}_*$ of thermalized regions in the BD eternal inflation model is determined by the two–dimensional probability flux of the inflaton and dilaton fields across the end–of–inflation boundary (43). This probability flux tells us the fraction of the volume of the universe that has undergone thermalization. Along the end–of–inflation boundary, the differential probability flux through a line element $\mathrm{d}l$ is given by $\mathrm{d}l \, (\vec{J} \cdot \hat{n})$, where $\hat{n}$ is a normal vector to the end–of–inflation curve (17). Hence we have

$$\mathcal{V}_* = \mathcal{V}_0 \left| \int_0^{t_c} \mathrm{d}t \, \mathrm{e}^{3t} \int_{EI} \mathrm{d}l \, (\vec{J} \cdot \hat{n}) \right|, \tag{85}$$

where the cutoff time $t_c$ is again defined by

$$\left| \int_{t_c}^{\infty} \mathrm{d}t \int_{EI} \mathrm{d}l \, (\vec{J} \cdot \hat{n}) \right| = \epsilon \left| \int_0^{\infty} \mathrm{d}t \int_{EI} \mathrm{d}l \, (\vec{J} \cdot \hat{n}) \right|. \tag{86}$$

For a given inflaton potential it is easy to show, after a little algebra, that the flux across the end–of–inflation boundary is given by

$$\int_{EI} \mathrm{d}l \, (\vec{J} \cdot \hat{n}) = \frac{C_0}{\pi} \mathrm{e}^{\gamma_1 t} \int \mathrm{d}\sigma \, \exp\left[8\pi\gamma_1 \, (3\omega - 2)\left(\frac{V}{V'}\right)^2\right] \left(\frac{V}{V'}\right) \left[2(3\omega - 2)\left(1 - \frac{VV''}{V'^2}\right) + 1\right]. \tag{87}$$

This expression may be computed either analytically or numerically depending on the potential, but a useful simplification can be introduced by taking the limit $\omega \to \infty$. The asymptotic form of (87) in this limit then reads

$$\int_{EI} dl \, (\vec{J} \cdot \hat{n}) \approx \frac{C_0}{8\pi^2 \gamma_1} \exp(\pi \gamma_1 \phi_{\max}^2 - \gamma_1 t), \tag{88}$$

where $\phi_{\max}$ is the maximum value the BD field can attain along the end–of–inflation boundary. In the case of a powerlaw potential this corresponds to $\sigma_* = 0$ and in the case of a double–well potential $\sigma_*^2 = \eta^2$. Therefore, in this limit (85) becomes

$$\mathcal{V}_* \approx \mathcal{V}_0 \frac{C_0}{8\pi^2 \gamma_1} \exp(\pi \gamma_1 \phi_{\max}^2) \frac{\exp\left[(3-\gamma_1)t_c\right]}{(3-\gamma_1)}, \tag{89}$$

and the cutoff time $t_c$ is determined by

$$t_c = -\frac{1}{\gamma_1} \log \epsilon. \tag{90}$$

Similarly, we now proceed to compute the regularized 3–volume $\mathcal{V}(\sigma, \phi)$ of the regions with arbitrary fields $(\sigma, \phi)$ still undergoing inflation. In order to do this, we note that the fraction of observers located at $(\sigma, \phi)$ within the lapse of time $t, t + dt$ is given by $|(\vec{J} \cdot \hat{l}) \, dt|$, where $\hat{l}$ is the tangent vector to the curve that crosses $(\sigma, \phi)$ given by the conservation law (20):

$$\phi^2 + 8 \int d\sigma \, \frac{V}{V'} = \text{const.} \tag{91}$$

Thus we have

$$\mathcal{V}(\sigma, \phi) = \mathcal{V}_0 \left| \int_0^{t_c} dt \, e^{3t} \, (\vec{J} \cdot \hat{l}) \right|, \tag{92}$$

and therefore from (66)(67)(91) we obtain

$$\mathcal{V}(\sigma, \phi) = \mathcal{V}_0 \frac{C_0}{4\pi} \left[\phi^2 + 16\left(\frac{V}{V'}\right)^2\right]^{1/2} \exp(\pi \gamma_1 \phi^2) \frac{\exp\left[(3-\gamma_1)t_c\right]}{(3-\gamma_1)}. \tag{93}$$

Hence, finally the ratio of the volume occupied by regions on a hypersurface $(\sigma, \phi)$ with respect to the thermalized regions in the $\omega \to \infty$ limit is then

$$r = \frac{\mathcal{V}(\sigma, \phi)}{\mathcal{V}_*} = 2\pi \gamma_1 \left[\phi^2 + 16\left(\frac{V}{V'}\right)^2\right]^{1/2} \exp\left[\pi \gamma_1 (\phi^2 - \phi_{\max}^2)\right], \tag{94}$$

or equivalently (via (20)),

$$r = 2\pi \gamma_1 \left[\phi_{\max}^2 - 8 \int d\sigma \frac{V}{V'} + 16\left(\frac{V}{V'}\right)^2\right]^{1/2} \exp\left(-8\pi \gamma_1 \int d\sigma \frac{V}{V'}\right). \tag{95}$$

Once again we note a tendency of the fields towards large values. For any inflaton potential the BD field is most likely to be close to $\phi_{\max}$. From the curve (91) this is equivalent to a tendency for small values of $\sigma$, but this is somewhat dependent on the specific shape of the potential. If $\phi_{\max}$ is not bounded, then the ratio $r$ is negligible for regions with finite values of the BD field. Hence, the largest physical volume is occupied by the hypersurface $\phi_{\max}$ and if we expect our observable universe to be a typical region of the universe, $\phi_{\max}$ is likely to be very large but finite, $\phi_{\max} \geq 10^{20}$ GeV. The regions where $\phi \approx \phi_{\max}$ are close to the end–of–inflation boundary, and therefore the largest fraction of the physical volume is occupied by inflating regions that are about to thermalize. For the inflaton field on the other hand, the form of $r$ depends entirely on the potential. In the case of a powerlaw potential there is again a tendency towards small values at the slope of the potential, in a trade–off between the competing factors $(V/V')$ and $e^{\pi\gamma_1\phi^2}$, whereas in the case of the double–well potential $r$ has a sharp peak centred at $\sigma = 0$ and it tails off at $\sigma^2 \approx \eta^2$. Only in the case of the exponential potential $e^{\alpha\sigma}$ is the ratio $r$ insensitive to the value of $\sigma$.

### 4.2.1 Powerlaw potential

In the case of the powerlaw potential, (87) can be calculated analytically. One obtains

$$\int_{EI} \mathrm{d}l\, \vec{J} \cdot \hat{n} = \frac{C_0}{(2\pi)^2\, n\gamma_1 \Delta} \left(1 + \frac{n}{2}\Delta\right) \exp(\pi\gamma_1 \Delta \sigma_0^2 - \gamma_1 t), \tag{96}$$

where

$$\Delta \equiv \frac{2}{n^2}(3\omega - 2), \tag{97}$$

and we have retained only the dominant contribution to the integral and used the explicit form of the probability current:

$$\vec{J} = (J_\sigma, J_\phi) = \frac{C_0}{4\pi}\left(\phi, -\frac{2}{n}\sigma\right) \exp\left(\pi\gamma_1\, \phi^2 - \gamma_1 t\right). \tag{98}$$

Therefore, bearing in mind that the time integration in (85) is dominated by its upper limit, one gets

$$\mathcal{V}_* \approx \mathcal{V}_0\, \frac{C_0}{(2\pi)^2\, n\gamma_1\Delta} \left(1 + \frac{n}{2}\Delta\right) \exp(\pi\gamma_1\Delta\sigma_0^2)\, \frac{\exp\left[(3-\gamma_1)t_c\right]}{(3-\gamma_1)}, \tag{99}$$

and similarly

$$\mathcal{V}(\sigma,\phi) \approx \mathcal{V}_0\, \frac{C_0}{4\pi}\left(\phi^2 + \frac{4}{n^2}\sigma^2\right)^{1/2} \exp(\pi\gamma_1\phi^2)\, \frac{\exp\left[(3-\gamma_1)t_c\right]}{(3-\gamma_1)}. \tag{100}$$

In this case, the conservation law employed tells us that (91) is an ellipse of the form $\phi^2 + (2/n)\sigma^2 = $ const. Thus, the volume ratio $r$ in this case is then

$$r = \pi\gamma_1\, n\Delta \left(1 + \frac{n}{2}\right)^{-1} \left(\phi^2 + \frac{4}{n^2}\sigma^2\right)^{1/2} \exp(\pi\gamma_1\phi^2 - \pi\gamma_1\Delta\sigma_0^2). \tag{101}$$

It is easy to see that this expression is consistent with (94) in the $\omega \to \infty$ limit. The value of $\sigma_0$ that dominates the integral (96) is the largest $\sigma$ accessible in the initial conditions. Considering that this value lies under the Planck boundary, it is consistent with the approximation $\phi_{\max}^2 \approx \Delta \sigma_0^2$ provided that the Planck boundary and the end–of–inflation boundary intersect at $\phi_{\max}$, i.e., $n > 2$. The alternative situation would be that of an unbounded $\phi_{\max}$ and in that case $\sigma_0$ would be unbounded too.

### 4.2.2 Double–well potential

In this case, the integral (87) can only be computed numerically, but to illustrate the results in an analytical form we use the asymptotic result (88) for the limit $\omega \to \infty$. Thus $\mathcal{V}_*$ is given by (89), and the volume $\mathcal{V}(\sigma, \phi)$ by (93). Hence, the ratio (94) is then

$$r = 2\pi \gamma_1 \left[\phi^2 + \frac{(\sigma^2 - \eta^2)^2}{\sigma^2}\right]^{1/2} \exp\left[\pi \gamma_1 (\phi^2 - \phi_{\max}^2)\right]. \tag{102}$$

As was noted in the previous section, this ratio peaks sharply at $\sigma \approx 0$ and tails off towards $\sigma^2 \approx \eta^2$. Therefore most regions are located in the neighbourhood of the top of the hill in the potential, as as we have seen in §2.2, this corresponds to the limit $\phi_{\max} \to \infty$. Provided $\phi_{\max}$ remains finite, then the conservation law (46) sets a lower bound for $\sigma^2$ and no regions exist where $\sigma$ is exactly zero, but in most regions it is situated in a small neighbourhood around this value.

## 5 Spectrum of density fluctuations in BD inflation

In this section, the spectrum of density fluctuations is calculated following the regularization scheme we have used above. We shall first derive the distribution of quantum fluctuations of the fields in BD eternal inflation that is measured by a 'typical' observer at a certain value of $H$. This distribution tells us the expectation value of the quantum fluctuations, $\langle \delta\sigma \rangle$ and $\langle \delta\phi \rangle$, that contribute to the classical field over a distance scale $H^{-1}$. Strictly speaking, the spectrum of density fluctuations is calculated via a Fourier decomposition of the stochastic fluctuations of the fields, such that only those contributions that extend over distance scales $H^{-1}$ contribute to the integral. Here we assume that the values of the fields are taken at the time when the perturbations enter the horizon and we investigate the fluctuations for a given value of $H$ for simplicity.

For an ensemble of observers located on a hypersurface $(\sigma_A, \phi_A)$, the fluctuations of the fields follow a Gaussian distribution due to the stochastic nature of inflation, with an additional factor that depends on the stage of inflation the fields are in, as discussed below. The spectrum of Gaussian fluctuations is described by the distribution

$$d\mathcal{P}_0(\delta\sigma, \delta\phi) = \frac{1}{(2\pi \Delta)^{1/2}} \exp\left(-\frac{\delta\sigma^2 + \delta\phi^2}{2\Delta^2}\right) d\delta\sigma \, d\delta\phi, \tag{103}$$

where we take the simple–minded view that the variance of the fluctuations is the same for both fields at first order in $H$, $\Delta \approx \frac{H}{2\pi}$ (Garcia–Bellido et al. 1994). The fluctuations around the hypersurface $(\sigma_A, \phi_A)$ depend on the physical volume occupied by the hypersurfaces in its neighbourhood, as it is likelier that the fields will jump from less probable values to more probable ones, and this likelihood is proportional to the ratio of the physical volumes occupied by each hypersurface. Thus the probability distribution observed by a typical observer at $(\sigma_A, \phi_A)$ is given by

$$d\mathcal{P}(\delta\sigma, \delta\phi) \sim \frac{\mathcal{V}(\sigma_A + \delta\sigma, \phi_A + \delta\phi)}{\mathcal{V}(\sigma_A, \phi_A)} \, d\mathcal{P}_0(\delta\sigma, \delta\phi), \tag{104}$$

and therefore the stationary value of this distribution with respect to the variations $\delta\sigma$ and $\delta\phi$ gives us the expectation value of the quantum jumps. These are

$$\langle \delta\sigma \rangle = \frac{16 \left(\frac{V}{V'}\right)\left(1 - \frac{VV''}{V'^2}\right) \Delta^2}{\left[\phi^2 + 16\left(\frac{V}{V'}\right)^2\right] - 16\left(1 - \frac{VV''}{V'^2}\right)\Delta^2}, \tag{105}$$

$$\langle \delta\phi \rangle = \frac{\Delta^2 \, \phi \left\{1 + 2\pi\gamma_1 \left[\phi^2 + 16\left(\frac{V}{V'}\right)^2\right]\right\}}{\left[\phi^2 + 16\left(\frac{V}{V'}\right)^2\right](2\pi\gamma_1\Delta^2 - 1) - \Delta^2}. \tag{106}$$

The resulting fluctuations are not Gaussian. These expressions can be simplified by the following approximations. As we have seen in the previous section, the predominant scenario is one where $\phi^2 \gg \sigma^2$, and the volume occupied by regions with large values of $\phi$ is by far the largest and therefore the most typical. Therefore

$$\langle \delta\sigma \rangle \approx \frac{16}{\phi^2} \left(\frac{V}{V'}\right)\left(1 - \frac{VV''}{V'^2}\right) \Delta^2, \tag{107}$$

$$\langle \delta\phi \rangle \approx \phi. \tag{108}$$

We note that the typical quantum jumps in the BD field get larger as the field grows in the course of inflation and, as derived in the volume ratio (94), regions with $\phi \approx \phi_{\max}$ occupy the largest volume. It is therefore expected that the typical value of the Planck mass at the end of inflation is that which corresponds to $\phi_{\max}$. The amplitude of the fluctuations of the inflaton field depend on the shape of the potential. In the case of the powerlaw potential we have

$$\langle \delta\sigma \rangle \approx \frac{\sigma}{n^2\pi^2} \frac{H^2}{\phi^2}, \tag{109}$$

and for the double–well potential,

$$\langle \delta\sigma \rangle \approx \frac{m^2\omega}{12\pi^2} \frac{1}{\sigma\phi^4} \left(1 + \frac{\eta^2}{\sigma^2}\right). \tag{110}$$

In the former case, the fluctuations are larger for high $\sigma$ and decrease in the course of inflation, and in the latter they are larger at small $\sigma$, and similarly they become smaller later as the field rolls down the potential towards $\pm\eta$. In all cases the quantum jumps of the inflaton field decrease for increasing $\phi$ and, as we have seen above, $\phi$ does increase in the course of inflation for generic potentials. Therefore, these two factors contribute to the non–Gaussianity of the fluctuations, and the end result is that typical quantum fluctuations are larger at the earlier stages of inflation, and the departure from these typical values is exponentially suppressed.

In order to compute the spectrum of density fluctuations given by the expectation values (107)(108) we use the standard result (see e.g. Linde 1990b)

$$\frac{\delta\rho}{\rho} = -\frac{6}{5} H \frac{\dot\sigma\,\delta\sigma + \dot\phi\,\delta\phi}{\dot\sigma^2 + \dot\phi^2}, \tag{111}$$

where the perturbations are given in the Einstein frame, $\tilde g_{\mu\nu} = \phi^2 g_{\mu\nu}$, and they enter the horizon during the matter–dominated era. Therefore, within the approximation (107)(108), this yields

$$\frac{\delta\rho}{\rho} = \frac{6}{5} H^2 \omega \frac{\frac{H^2}{\pi^2\phi}\left(1 - \frac{VV''}{V'^2}\right) - \phi}{H^2\phi + \frac{\phi^3}{4}H'^2}, \tag{112}$$

which in the limit of large $\phi$ is reduced to

$$\frac{\delta\rho}{\rho} = \frac{48\pi}{5}\left(\frac{H}{H'}\right)^2 \frac{1}{M_P^2(\phi_{\max})}. \tag{113}$$

For an arbitrary potential it is easy to calculate the spectrum of fluctuations given by the superposition of all the individual contributions (113) of each mode. It can be safely assumed that, given that the BD field is almost constant after the end of inflation and its typical value is $\sim \phi_{\max}$, then typically $\phi_* \sim \phi_{\max}$. On the other hand, as shown in §2, $H/H'$ at the end of inflation is typically one or two orders of magnitude smaller than the Planck constant. This is however not satisfactory to make (113) sufficiently small so that, in agreement with observational constraints, $\langle\delta\rho/\rho\rangle \lesssim 10^{-4}$, and we require that the inflaton field be much smaller than its typical value at the end of inflation. For the case of the powerlaw potential, $\sigma_*/n \sim 10^{18}$ GeV, and thus one would obtain a density contrast $\langle\delta\rho/\rho\rangle \sim 10^{-4}$ only if predominantly $\sigma \sim 10^{-2}\sigma_*$.

In those cases where $\phi_{\max}$ is unbounded, most of the volume of the universe will be occupied by a smooth distribution of matter and the Planck mass will grow without limit. In such scenarios, regions where the Planck mass at the end of inflation is large but finite are untypical, and emerge from quantum jumps across the end–of–inflation boundary that prevent $\phi$ from taking arbitrarily large values along the classical trajectory. This behaviour does not arise in the case for powerlaw potentials $n > 2$, for which the end–of–inflation and Planck boundaries cross; these potentials permit us to predict a finite value of $\phi_{\max}$ that tells us the typical value of the Planck mass at the end of inflation.

# 6    Conclusions

In this paper we have investigated the probability distributions and spectrum of density fluctuations in a cosmological scenario of BD theories with eternal inflation. We have especially focused on powerlaw and double–well inflaton potentials, giving explicit results in both particular cases. In the calculation of the probability distributions, we have studied the solutions of the diffusion equation in the slow–roll regime and used these results to compute the ratio of physical volumes of hypersurfaces with arbitrary values of the fields with respect to thermalized regions. In accordance to the principle of global stationarity, these 3–volume ratios tell us the fraction of physical volume occupied by arbitrary values of the fields, in a time–independent manner, at any stage of inflation.

The physical volumes of the hypersurfaces considered are naturally divergent, and hence we have adopted a regularization procedure suggested by Vilenkin (1995a) to compute the ratios of any two volumes. Recently it has been shown (Linde & Mezhlumian 1996) that different regularization procedures can be chosen, that enable us to compute volume ratios in a way that is independent of the choice of time parametrization, but which yield mutually incompatible results. Previous work on the calculation of the spectrum of density fluctuations in BD inflation (see e.g. Garcia–Bellido *et al.* 1994) shows a strong dependence of the results on the choice of time parametrization. On the other hand, the approach undertaken here is dependent on the choice of the regularization procedure. This is a fundamental issue to be addressed, but I believe that we are moving on the right direction.

As shown by Winitzki & Vilenkin (1996), the residual mild dependence on the choice of time parametrization in the regularization procedure of Vilenkin (1995a) is due to the no–diffusion approximation employed in solving the stochastic equations of motion for $\mathcal{P}$. It is shown in the same paper that the error brought in by this approximation is of the same order of magnitude as the "diffusion terms" neglected (first term on the RHS of (55)). In principle, a perturbative approach is valid to solve the system up to a given order and restrict the dependence on the time parametrization to arbitrary accuracy. At the same time, this perturbative approach would miminize the dependence of the results on the regularization procedure, since both effects are closely related. Ideally an exact analytical solution of the equations would permit us to calculate the spectrum of fluctuations in a way that is fairly independent of the choice of regularization employed.

In §2 we have derived a useful conservation law, (20), that helps us to extract some information of the final configurations in terms of the initial conditions. In particular, we conclude from this conservation law that, for the case of a double–well potential, a scenario where $\sigma_0 \approx 0$ is only compatible with $\phi_{\max} \to \infty$. In this case, like in powerlaw potentials $n \leq 2$, we find that 'run–away' solutions take place, i.e. the leading contribution to the volume of the universe is given by configurations where the fields grow indefinitely. In the probability distributions we have calculated, it is found that in a typical region the BD field has a value that is close to $\phi_{\max}$. If $\phi_{\max}$ is unbounded, most regions will be totally smooth and cosmic structure will not form, $\langle\delta\rho/\rho\rangle \sim 0$, and thus it will be most unlikely to find regions where $\phi_*$ is finite. However, such scenarios are not entirely ruled out, but would lead us to the conclusion that, if this is a plausible scenario in our universe, then we inhabit a rare and highly untypical region. If on the other hand our region of the universe is typical, then a realistic inflaton potential ought to yield a finite $\phi_{\max}$. If at the initial time $\phi_0 \gg \sigma_0$, $\phi_{\max}$ is close to $\phi_0$ and therefore its magnitude becomes a particle physics problem of initial conditions.

The conservation law (20) also permits us to write all quantities in the classical evolution in terms of one field only. The probability distributions (82)(83)(84) may be written in terms of the inflaton field only with the aid of (27)(46), as well as the volume ratios, as given by (95). Therefore, one can derive the probability distribution for the density field during inflation, $V \sim \rho$. For instance, in the case of the powerlaw potential, this is such that the distribution has a stationary point at $\rho \approx \left(n\lambda/32\pi^2\gamma_1^2\right)^{1/2n}$. For a very approximate and conservative estimate of the parameters, such as $n \approx 2$, $\gamma_1 \approx 3$, this yields $\rho \approx 0.16\,\lambda^{0.25}$, which a reasonable typical value to seed galaxy formation in our own universe.

In the study of the typical quantum jumps in §5 one takes into account that both fields evolve independently of one another and therefore (20) is not satisfied in that regime. The typical quantum jumps of the fields during the course of inflation do not follow a Gaussian distribution due to biased effect of the volume ratios of the hypersurfaces between which the quantum transition takes place. This ratio represents the likelihood of a quantum jump in terms of the fraction of physical space occupied by each configuration. The typical quantum jumps take a simple form in the limit of large $\phi$, which is, as we have seen from the the probability distributions, the likeliest scenario. It is found, as expected, that the amplitude of the fluctuations decreases in the course of inflation and it is also smaller for larger $\phi$. In those regions where $\phi$ grows indefinitely, one can to a good approximation consider that the classical evolution is valid throughout and quantum jumps are negligible. The spectrum of density fluctuations obtained in this scenario is expressed in terms of the relevant parameters of the potential and $\phi_{\max}$, and one is therefore able to test models so that, provided $\phi_{\max}$ remains finite, the values of the parameters are consistent with the astrophysical bounds for $\langle\delta\rho/\rho\rangle$.

# Acknowledgments

The author would like to thank Andrei Linde and Juan Garcia–Bellido for useful comments on an earlier version of the manuscript.

# References

[1] Accetta F.S., Krauss L.M., Romanelli P., 1990, Phys. Lett. B 248, 146

[2] Albrecht A., 1995, in proc. of *The Birth of the Universe*, ed. Occhionero F., Springer–Verlag, Berlin

[3] Barrow J.D., Maeda K., 1991, Nucl. Phys. B341, 294

[4] Barrow J.D., Tipler F.J., 1986, The Anthropic Cosmological Principle, Clarendon Press, Oxford

[5] Bergmann P.G., 1968, Int. J. Theor. Phys. 1, 25

[6] Brans C.H., 1962, Phys. Rev. 125, 2194

[7] Brans C.H., Dicke R.H., 1961, Phys. Rev. 124, 925

[8] Dicke R.H., 1962, Phys. Rev. 125, 2163

[9] Carter B., 1983, Philos. Trans. R. Soc. London, A310, 347

[10] Casas J.A., Garcia-Bellido J., Quirós M., 1992, Phys. Lett. B278, 94

[11] Coleman S., 1988, Nucl. Phys. B307, 867

[12] Damour T., Nordtvedt K., 1993, Phys. Rev. D48, 3436

[13] Garcia–Bellido J., 1994, Nucl. Phys. B 423, 221

[14] Garcia–Bellido J., Linde A.D., 1995, Phys. Rev. D51, 429

[15] Garcia–Bellido J., Linde A.D., Linde D.A., 1994, Phys. Rev. D50, 730

[16] Goncharov A.S., Linde A.D., Mukhanov V.F., 1987, Int. J. Mod. Phys. A2, 561

[17] Hawking S.W., 1987, in *300 years of gravitation*, eds. Hawking S.W., Israel W., Cambridge University Press, Cambridge

[18] Hartle J., Hawking S.W., 1983, Phys. Rev. D28, 2960;

[19] La D., Steinhardt P.J., 1989, Phys. Rev. Lett. 62, 376

[20] Laycock A.M., Liddle A.R., 1994, Phys. Rev. D49, 1827

[21] Liddle A.R., Wands D., 1992, Phys. Rev. D45, 2665

[22] Lidsey J.E., 1996, Class. Quantum Grav., to appear

[23] Linde A.D., 1984, Zh. Eksp. Teor. Fiz. 87, 369 [Sov. Phys. JETP 60, 211]

[24] Linde A.D., 1986, Phys. Lett. B175, 395

[25] Linde A.D., 1987, Physica Scripta T15, 169

[26] Linde A.D., 1989, Phys. Lett. B227, 352

[27] Linde A.D., 1990a, Phys. Lett. B238, 160

[28] Linde A.D., 1990b, 'Particle Physics and Inflationary Cosmology', Harwood Academic, Switzerland

[29] Linde A.D., 1994a, Phys. Rev. D49, 748

[30] Linde A.D., 1994b, Lectures on Inflationary Cosmology, hep–th/9410082

[31] Linde A.D., 1995, preprint gr–qc/9508019

[32] Linde A.D., Mezhlumian A., 1996, Phys. Rev. D53, 4267

[33] Linde A.D., Mezhlumian A., 1993, Phys. Lett. B307, 25

[34] Linde A.D., Linde D.A., Mezhlumian A., 1994, Phys. Rev. D49, 1783

[35] Linde A.D., Linde D.A., Mezhlumian A., 1995, Phys. Lett. B345, 203

[36] Mijić M., 1990, Phys. Rev. D42, 2469

[37] Mijić M., 1991, Int. J. Mod. Phys. A6, 2685

[38] Nordtvedt K., 1970, Ap. J. 161, 1059

[39] Reasenberg R.D., *et al.* 1979, Astrophys. J. 234, L219

[40] Rees M.J., 1983, Philos. Trans. R. Soc. London A310, 311

[41] Starobinsky A.A., 1986, in *Field Theory, Quantum Gravity and Strings*, eds. de Vega H.J., Sanchez N., Springer–Verlag, Heidelberg, 206, p.107

[42] Starobinsky A.A., 1984, in *Fundamental Interactions*, MGPI Press, Moscow, p.55

[43] Vilenkin A., 1983, Phys. Rev. D27, 2848

[44] Vilenkin A., 1984, Phys. Rev. D30, 549

[45] Vilenkin A., 1995a, Phys. Rev. D52, 3365

[46] Vilenkin A., 1995b, Phys. Rev. Lett. 74, 846

[47] Wagoner R.V., 1970, Phys. Rev. D1, 3209

[48] Will C.M., 1993, Theory and Experiment in Gravitational Physics, 2nd ed., Cambridge University Press, Cambridge

[49] Winitzki S., Vilenkin A., 1996, Phys. Rev. D53, 4298

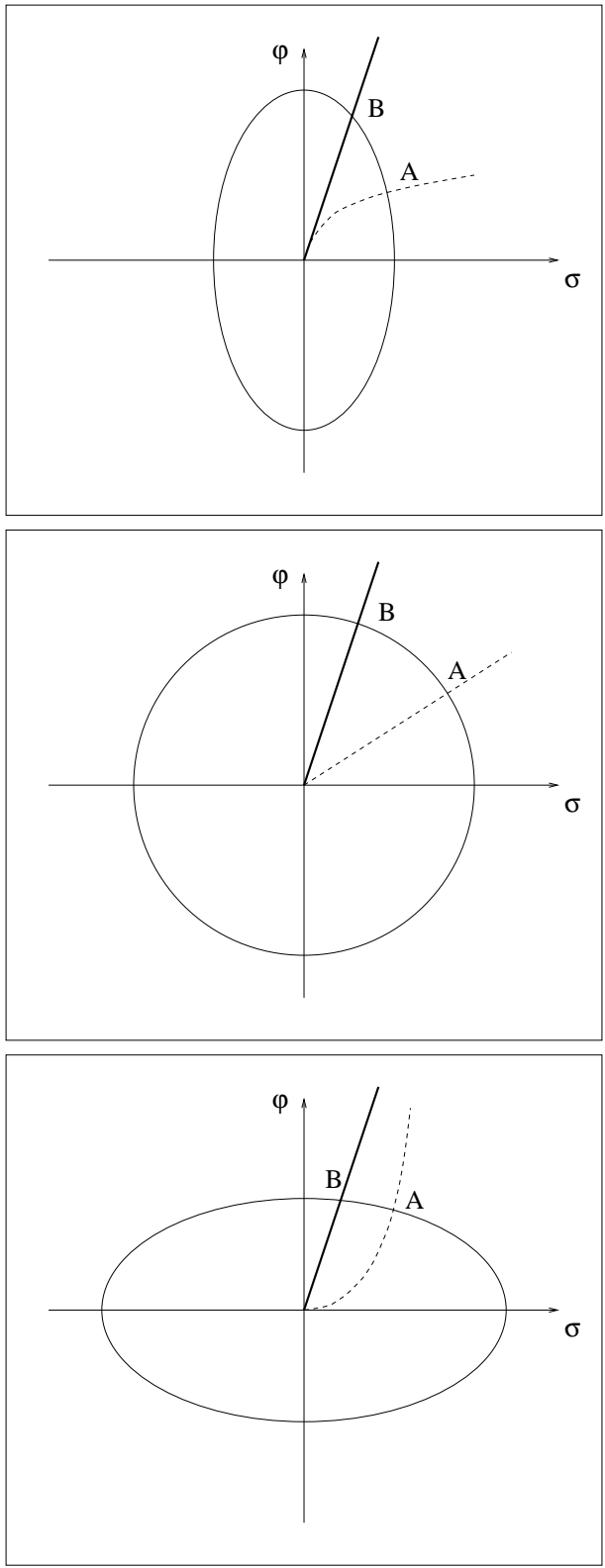

Figure 1: Field trajectories on the $(\sigma, \phi)$ plane for a power–law potential. As discussed in §2.1, the classical trajectories are ellipses, their axes determined by the initial conditions and $n$. The dashed line is the Planck boundary and the thick solid line is the end–of–inflation boundary. For an initial condition located at $A$, inflation will occur for as long as the fields roll down the potential until they reach $B$. *Top*: $n < 2$; *centre*: $n = 2$; *bottom*: $n > 2$.

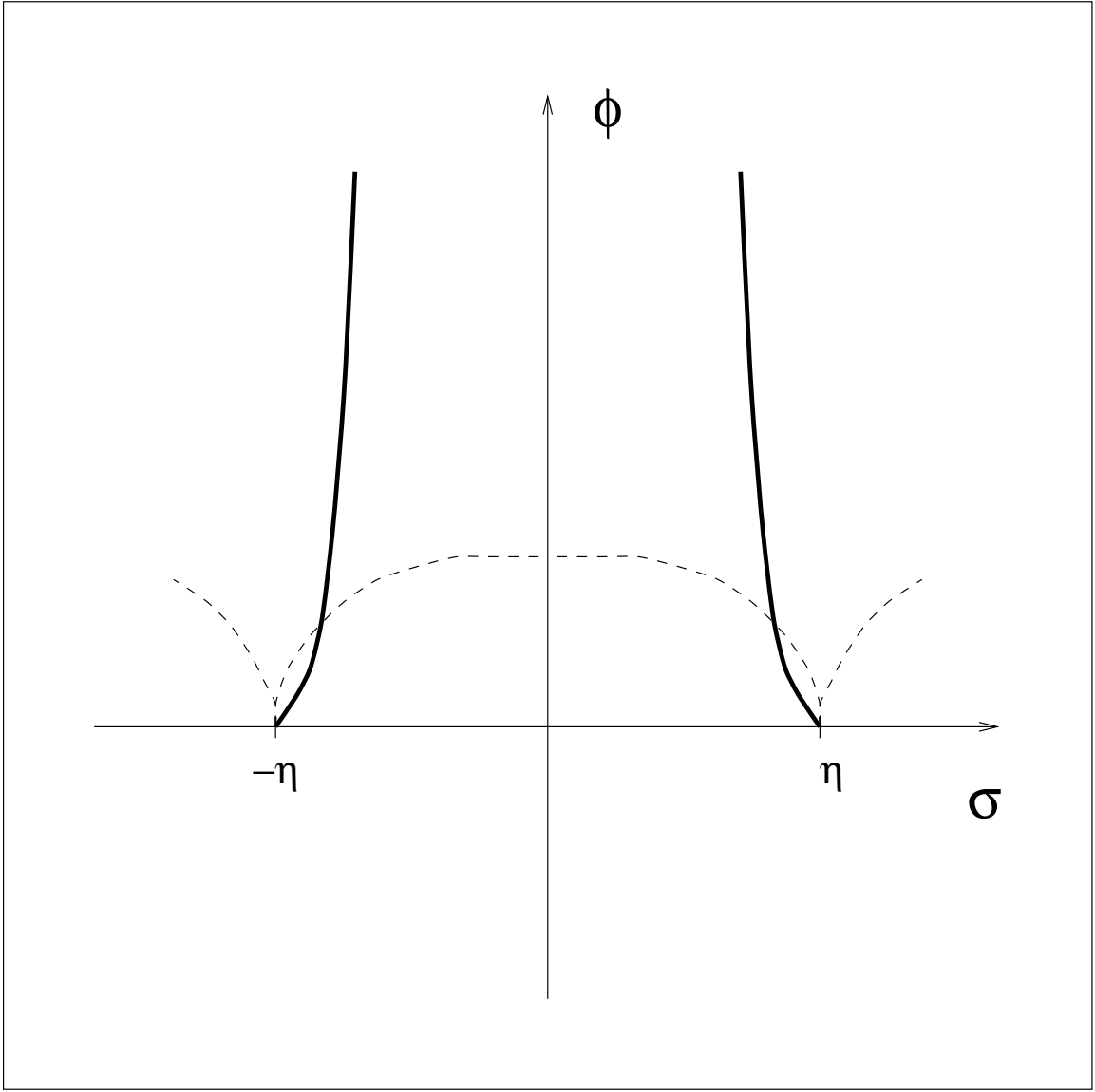

Figure 2: Planck boundary (dashed line) and end–of–inflation boundary (thick full line) for the double–well potential of §2.2. The allowed values of the fields lie in the region above the Planck boundary and inflation only takes place for values below the end–of–inflation boundary. It must be noted that there is a small neighbourhood of $\sigma^2 \approx \eta^2$ that is never reached as a result of inflation but becomes negligibly small in the limit $\eta \to \infty$.